\documentclass[11pt,twocolumn,twoside,a4paper,amsmath,amssymb,aps,showkeys,showpacs]{revtex4}

\usepackage{amsfonts}
\usepackage{graphics} %   <------- for LatTeX + EPS
\usepackage{epsfig}
\usepackage{fancyheadings}

\newcommand{\beq}{\begin{equation}}
\newcommand{\eeq}[1]{\label{#1} \end{equation}}

\begin{document}
\thispagestyle{myheadings}
%%%%%%%%%%%%%%%%%%%%%%%%%% Title %%%%%%%%%%%%%%%%%%%%%%%%%%%%%%%%%%%%%%
\rhead[]{}%<------
\lhead[]{}%<------
\chead[V.K. Magas, J. Yamagata-Sekihara, S. Hirenzaki, E. Oset, A. Ramos]{The (K-,p) reaction on C12 at KEK}%<------short title

\title{The ($K^-,p$) reaction on $ ^{12}C$ at KEK}

\author{V.K. Magas}
\affiliation{Departament d'Estructura i Constituents de la Materia, Universitat de Barcelona, Diagonal 647, 08028 Barcelona, Spain}
%\email{vladimir@ecm.ub.es}

\author{J. Yamagata-Sekihara}
\affiliation{Yukawa Institute for Theoretical Physics, Kyoto University, Kyoto 606-8502, Japan}
%\email{yamagata@yukawa.kyoto-u.ac.jp}

\author{S. Hirenzaki}
\affiliation{Department of Physics, Nara Women's University, Nara 630-8506, Japan}
%\email{zaki@cc.nara-wu.ac.jp}

\author{E. Oset}
\affiliation{Dep. de F\'{\i}sica Te\'orica and IFIC, Centro Mixto Universidad de Valencia-CSIC,
Institutos de Investigaci\'on de Paterna, Apartado 22085, 46071 Valencia, Spain}
%\email{oset@ific.uv.es}

\author{A. Ramos}
\affiliation{Departament d'Estructura i Constituents de la Materia, Universitat de Barcelona, Diagonal 647, 08028 Barcelona, Spain}
%\email{ramos@ecm.ub.es}

%\received{ ?????????? }

\begin{abstract}
We study the $(K^-,p)$ reaction on $ ^{12}C$ with a kaon beam of $1$ GeV momentum, paying a special attention to the region of emitted protons having kinetic energy above 600 MeV, which was used to claim a deep kaon nucleus optical potential \cite{Kishimoto:2007zz}. 
The experiment looks for fast protons emitted from the absorption of in flight kaons by nuclei, but in coincidence with at least one charged particle in the decay counters sandwiching the target. The analysis of the data is done in \cite{Kishimoto:2007zz} assuming that the coincidence requirement does not change the shape of the final spectra. However our detailed calculations show that this assumption doesn't hold, and, thus, the final conclusion of this experiment is doubtful. 

We perform Monte Carlo simulation of this reaction. The advantage of our method with respect to Green's function method used in \cite{Kishimoto:2007zz} is that it   
allows to account not only for quasi-elastic $K^- p$ scattering, but also for  the other processes which contribute to the proton spectra. 
We investigated the effect  of the multi-scatterings and of the $K^-$ absorptions by one and two nucleons ($K^- N \to \pi Y$ and $K^- N N\to  Y N$) followed by the decay of the hyperon in $\pi N$. We show that all these mechanisms allow us to explain reasonably well the observed spectrum with standard shallow kaon nucleus optical potential, obtained in chiral models.   
\end{abstract}



\maketitle

%\renewcommand{\thefootnote}{\fnsymbol{footnote}}
%{\footnotetext[1]{Also delivered by SGn as evening (after dinner)
%talk at the ITI Conference in Bad Zwischenahn, October 2005}

\renewcommand{\thefootnote}{\roman{footnote}}

%*****************   The Body of the Article:   *************************

%\section{Introduction}
%\label{introduction}

    The issue of the kaon interaction in the nucleus has attracted much
attention in past years. Although from the study of kaon atoms one knows that
the $K^-$-nucleus potential is attractive \cite{friedman-gal}, the discussion
centers on how attractive the potential is and whether it can accommodate deeply
bound kaon atoms (kaonic nuclei), which could be observed in direct reactions.

All modern potentials
based on underlying chiral dynamics of the $KN$ interaction
\cite{lutz,angelsself,schaffner,galself,Tolos:2006ny} lead to
moderate potentials of the order of 60 MeV attraction at normal nuclear density.
They also have a large imaginary part making the width of the bound states
much larger than the energy separation between the levels, which would rule out
the  experimental observation of these states.

Deep  $K^-$-N optical potentials are preferred by the phenomenological fits to kaon atoms data. 
One of the most known extreme cases of these type is a highly
attractive phenomenological potential with about 600 MeV strength in the center of the nucleus, 
introduced in \cite{akaishi:2002bg,akainew}.  
In these picture  such an attractive $K^-$, inserted inside the nucleus,  would lead to a shrinkage of the nucleus, generating a new very compact object - kaonic nucleus - with central density which can be 10 times larger than normal nuclear density. Such super-deep
potentials were criticized in \cite{toki,Hyodo:2007jq,Oset:2007vu,npangels}.

It is important to keep in mind that in kaon atoms the $K^-$ is primary bound by the Coulomb force. These are extended systems, and therefore their properties
cannot directly tell us about the $K^-$-N potential at short distances.  From the experimental side the search for bound $K^-$ states with nucleons is a most direct and clear way to answer whether the $K^-$-nucleon potential is deep or shallow, because only deep potential may generate states sufficiently narrow to be observed experimentally. Experimental attempts to resolve this situation are made since 2004, but the situation is still very unclear. 

Several claims of observed deeply bound $K^-$ states have been made. However the first one, about $K^-pnn$ state bound by 195 MeV from the 
experiment at KEK \cite{Suzuki:2004ep}, is now withdrawn after a new more precise experiment \cite{Sato:2007sb}.
The peaks seen by FINUDA  and originally
interpreted in terms of deeply bound $K^-pp$ \cite{Agnello:2005qj} and $K^-ppn$  \cite{:2007ph} clusters, are now put under 
the question, because in Refs. 
 \cite{Magas:2006fn,Ramos:2007zz,Crimea,Magas:2008bp} these peaks found 
 explanations based on conventional reactions that unavoidably occur in the process of
 kaon absorption. 
 
There are also claims (with very low statistical significance) of $K^-pp$ and $K^-ppn$ bound states from $\bar{p}$ annihilation in $ ^4He$ at rest measured by OBELIX@CERN \cite{Obelix}. And the recent claim of $K^-pp$ bound state, seen in $pp\rightarrow K^+ X$  reaction, from DISTO experiment \cite{DISTO}. These experimental claims are under investigation now.  Before calling in new physics one has to make sure that these data cannot be explained with conventional mechanisms. 
 
There is however one more experiment where the authors claim
the evidence for a strong kaon-nucleons potential, with a depth
of the order of 200 MeV \cite{Kishimoto:2007zz}.  
The experiment
looks for fast protons
emitted from the absorption of in flight kaons by $^{12}C$ in coincidence with at least one charge particle in the decay counters sandwiching the target.
The data analysis in \cite{Kishimoto:2007zz} is based on the assumption that the coincidence requirement does not change the shape of the final spectra.
We shall see that this assumption doesn't hold and the
interpretation of the data requires a more thorough approach than the one
used in  that work. 

One of the shortcomings of Ref.~\cite{Kishimoto:2007zz} stems from employing the
Green's function method \cite{Morimatsu:1994sx} to analyze the data.  
The only mechanism considered in
Ref.~\cite{Kishimoto:2007zz} for the emission of fast protons is the $\bar{K} p \to
\bar{K} p$ process, taking into account the optical potential for the slow
kaon in the final state. 
We shall show that there are other mechanisms that
contribute to generate fast protons, namely multi-scattering reactions, and kaon
absorption by one nucleon, $K^- N \to \pi \Sigma$ or $K^- N \to \pi \Lambda$
or by a pair of nucleons, $\bar{K} N N\to  \Sigma N$
and $\bar{K} N N\to  \Lambda N$, followed by decay of $\Sigma$ or $\Lambda$ into $\pi N$.
The contributions from these processes were also suggested in Ref. \cite{YH}.

%\section{Monte Carlo simulation}
In the present work, we take
into account all the above mentioned reactions by means
 of a Monte Carlo simulation \cite{simulation}.
The election of which reaction occurs at a certain point in the nucleus
is done as usual. One chooses a step size $\delta l$ and calculates, by means of $\sigma_i \rho \delta l$, the probabilities that any of
the possible reactions happens $i=$ {\it Quasi-elastic, 1N absorption, 2N absorption}; $\rho$ is nucleon density.
 The size of $\delta l$ is
small enough such that
the sum of probabilities that any reaction occurs is reasonably smaller
than unity. A random number from 0 to 1 is generated and a reaction occurs if
the number falls within the corresponding segment of length given by its probability.
If the random number falls outside the sum of
all segments then
this means that no reaction has taken place and the kaon is allowed to proceed one further step $\delta l$.
The simulation of one event is over when all the produced particles leave the nucleus.  
To adapt the calculations to the experiment of \cite{Kishimoto:2007zz}  we select "good events" with fast protons  that emerge
within an angle of 4.1 degrees in the nuclear rest frame (lab frame). 
As in \cite{Kishimoto:2007zz} we plot our obtained $^{12}$C$(K^-,p)$ spectrum as a function of a binding energy of the kaon, $E_B$, should
the process correspond to the trapping of a kaon in a bound state and emission of the fast proton.
%, according to
%$
%\sqrt{(E_K+M_{^{12}C}-E_p)^2-(\vec{P}_p - \vec{P}_K)^2} 
%= M_{^{11}B} + M_{K}
%-E_B \,,
%$
%where $E_p,\vec{P}_p$ are the energy and momentum of the observed proton and
%$E_K,\vec{P}_K$ are the energy and momentum of the initial kaon.

%{\bf Quasi-elastic scattering.} \ \ \
If there is a 
quasi-elastic collision at a certain point, then
the momentum of the $K^-$ and that of the nucleon, which is
randomly chosen within the Fermi sea, are boosted to their CM frame. 
The direction of the scattered momenta is determined according to the 
experimental cross section.
A boost to the lab frame determines the final kaon and nucleon momenta. The event is kept as long as the size of the nucleon momentum is larger than the local Fermi momentum.
Since we take into account secondary collisions we also
consider the reaction $K^- p \to K^0 n$ and $K^- n \to K^- n$ with their
corresponding cross sections.

Once primary nucleons are produced they are also followed through the nucleus taking into account the probability that they collide with other nucleons, losing
energy and changing their direction, see \cite{Magas:2006fn,Ramos:2007zz,Crimea,Magas:2008bp} for more details.

We also follow the rescattered kaon on its way through the nucleus. In the subsequent interaction process we let the kaon experience
whichever reaction of the three that we consider (quasi-elastic, one-body absorption, two-body absorption) according to
their probabilities. This procedure continues until kaon is absorbed or leaves of the nucleus.

Apart from following the kaons and nucleons, our calculations also need to consider
the quasi-elastic scattering of $\Lambda$'s and
$\Sigma$'s (produced in the kaon absorption reactions) on their way through the residual nucleus. 
Given the uncertainties in the hyperon-nucleon cross sections, we 
may use for $\Sigma N$ scattering the  relation $\sigma_{\Sigma N} = 2\sigma_{NN}/3$, based on a simple
non-strange quark counting rule. In the case of $\Lambda N$ scattering, we use the
refined parameterization of Ref.~\cite{manolo},  as was also done in Ref.~\cite{Crimea}.

%{\bf One and two body kaon absorption.}\ \ \ 
One nucleon $K^-$ absorption leads to $K^- N$ $\to \pi \Lambda$ or $K^- N \to \pi \Sigma$,
with all the possible charge combinations.
The elastic and inelastic two-body ${\bar K}N$ 
cross sections for kaons are taken from the Particle Data Group \cite{PDG}.

The kaon absorption by two nucleons is a bit more tricky. 
Here we take into account the following
processes: $K^- NN \to \Lambda N$ or $K^- NN \to \Sigma N$ with all possible
charge combinations. 
In these
reactions an energetic nucleon is produced, as well as a $\Lambda$ or a
$\Sigma$. Both the nucleon and the hyperon are followed through the nucleus as
discussed above. 
Once out of the nucleus, the hyperons are let to decay weakly
into $\pi N$ pairs. 
Therefore, the two-body absorption process provides a double source of
fast protons, those directly produced in two nucleon absorption reaction
and those coming from hyperon decays.

We assume a total two body absorption rate to be 20\% that of one body absorption at
about nuclear matter density, something that one can infer from
data of $K^-$ absorption in $^4$He  \cite{Katz:1970ng}.
In practice, this is implemented in the following way.
The probability per unit length for two nucleon absorption is proportional to the square of the nucleon density:
$\mu_{K^-NN}(\rho) = C_{\rm abs} \rho^2\,.$
We assume that
$ \langle \mu_{K^-NN}\rangle = C_{\rm abs} \langle \rho^2 \rangle = 
0.2 \langle \mu_{K^-N} \rangle = 0.2 \sigma^{\rm tot}_{K^-N} \langle \rho \rangle \,,$
where $\sigma^{\rm tot}_{K^-N}$ accounts for the total one nucleon absorption cross section and, in symmetric nuclear matter it is given by:
$\sigma^{\rm tot}_{K^-N}=(\sigma^{\rm tot}_{K^-p} + \sigma^{\rm tot}_{K^-n} - 
\sigma_{K^-p\rightarrow K^-p} - \sigma_{K^-n\rightarrow K^-n})/2\,.  $
Taking $\langle \rho \rangle=\rho_0/2$, where $\rho_0=0.17$ fm$^{-3}$ is normal nuclear matter density, we obtain
$C_{\rm abs}\approx 6\ {\rm fm}^{5} \,.$

The different partial processes that can take place in a two-nucleon 
absorption reaction are:
$K^- p p \to p \Lambda,\ p \Sigma^0,\ n \Sigma^+ $; 
$K^- p n \to n \Lambda,\ n \Sigma^0,\ p \Sigma^- $; 
$K^- n n \to n \Sigma^- \ .$
Ideally, their corresponding branching ratios should be obtained from relevant
microscopic mechanisms, however in the present exploratory work, we will
consider a much simpler approach consisting of assigning equal probability to
each of the above reactions. Noting that the chance of the kaon to find a
$pn$ pair is twice as large as that for $pp$ or $nn$ pairs, we 
finally assign a probability of 3/10 for having a $p\Sigma$ pair in the final
state, 4/10 for $n \Sigma$, 1/10 for $p \Lambda$ and 2/10 for $n \Lambda$.

\begin{figure}[htb]
\vspace{-0.25cm}
\includegraphics[width=.5\textwidth]{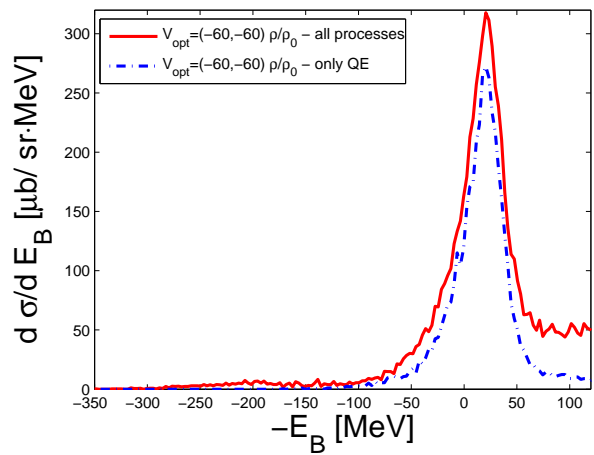}
\vspace{-0.75cm}
\caption{Calculated $ ^{12}C(K^-,p)$ spectra 
 with  $V_{\rm opt}=(-60,-60)\rho/\rho_0$ MeV,
  taking into account only quasi-elastic processes (dash-dotted line),  
  and including all the contributing processes (full line).}
\label{fig1}
\end{figure}

%{\bf Implementation of the $K^-$ optical potential.} \ \ \
We also take into account a kaon optical potential $V_{\rm opt}={\rm Re}\, V_{\rm
opt} + {\rm i}~ {\rm Im}\, V_{\rm opt} $, which will influence the kaon propagation
through the nucleus, especially when it will acquire a relatively low momentum 
after a high momentum transfer quasi-elastic collision.  
In the present study we take the strength of the potential as predicted by chiral models:
${\rm Re}\, V_{\rm opt}= -60\, \rho/\rho_0$ MeV \cite{lutz,angelsself,schaffner,galself,Tolos:2006ny}; 
${\rm Im}\, V_{\rm opt} \approx -60\, \rho/\rho_0$ 
MeV, as in
the experimental paper \cite{Kishimoto:2007zz} and the theoretical study of
\cite{angelsself}.

In the Monte Carlo simulation we implement this distribution by
generating a random kaon mass $\tilde{M}_K$ around a central value, 
$M_K + {\rm Re}\,V_{\rm opt}$, within 
a certain extension determined by the width of the distribution
$\Gamma_K = -2 {\rm Im}\, V_{\rm opt}$. The probability assigned to each value
of $\tilde{M}_K$ follows the Breit-Wigner distribution given by the kaon 
spectral function:
\begin{center}
$
S(\tilde{M}_K )=\frac{1}{\pi} 
\frac{-2M_K {\rm Im}\,V_{\rm opt}}
{(\tilde{M}^2_K -M_K^2-2 M_K {\rm Re}\,V_{\rm opt})^2 + (2M_K {\rm Im}\,V_{\rm opt})^2}
\,.
$
\end{center}

%{\bf Results and discussion.}\ \ \
In Fig.~\ref{fig1}
we show  the results
of the Monte Carlo simulation obtained with an optical potential
$V_{\rm opt}=(-60,-60)\rho/\rho_0$ MeV:  first, taking into account only quasi-elastic
processes; and then taking into account all the discussed mechanisms. We can see that there is some 
strength gained in the region of "bound kaons" due to the new mechanisms.
Although not shown separately in the figure, we have observed
that one nucleon absorption and several rescatterings
contribute to the region $-E_B > -50$ MeV. To some extent, this strength
can be simulated by the parametric background used in
\cite{Kishimoto:2007zz}. However, this is not true anymore for the two nucleon absorption process,
which contributes to all values of $-E_B$, starting from almost as low as $-300$ MeV.

It is very important to keep in mind that in the spectrum of \cite{Kishimoto:2007zz} 
the outgoing forward protons were measured in  coincidence  
with at least one charged particle in the decay counters sandwiching the target.
Obviously, the real simulation of such a coincidence experiment is tremendously
difficult, practically impossible with high accuracy, because it
would require tracing out all the charged particles coming out from all
possible scatterings and decays. 
Although we are studying many processes and 
following many particles in our
Monte Carlo simulation, which is not the case in the Green function method
used in the data analysis \cite{Kishimoto:2007zz}, 
we cannot simulate precisely the real coincidence effect.

\begin{figure}[htb]
\vspace{-0.25cm}
\includegraphics[width=.5\textwidth]{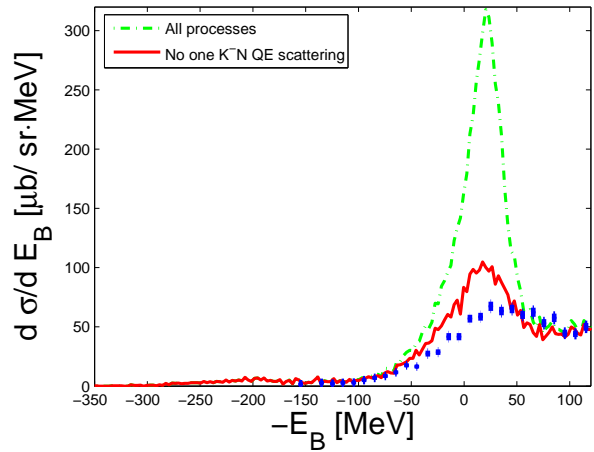}
\vspace{-0.75cm}
\caption{Calculated $ ^{12}C(K^-,p)$ spectra 
 with  $V_{\rm opt}=(-60,-60)\rho/\rho_0$ MeV
  taking into account all contributing processes (dash-dotted line). Then we impose minimal 
  coincidence requirement (full line). Data points are from  \cite{Kishimoto:2007zz}.}
\label{fig2}
\end{figure}

The best we can do is to eliminate the processes
which, for sure, will not produce a coincidence, this can be 
called minimal coincidence requirement.
If the kaon in the first
quasi-elastic scattering produces an energetic proton falling into the
peaked region of the spectra, then the emerging kaon will be 
scattered backwards.
In our Monte Carlo simulations we can select events were neither the proton,
nor the kaon will have any further reaction after such a scattering. In these
cases, although there is a "good" outgoing proton, there are no charged
particles going out with the right angle with respect to the beam axis to
hit a decay counter, since the $K^-$ escapes undetected in the backward
direction. Therefore, this type of events must be eliminated for 
comparison with the experimental spectra.

It is clear from Fig.~\ref{fig1} that the main source of the energetic protons for $ ^{12}C(K^-,p)$ spectra is
$K^-p$ quasi-elastic scattering, however many of these events will not pass the coincidence condition.
Implementing the minimal coincidence requirement, as discussed above, 
we will cut off a substantial part of the potentially "good" events, and drastically
change the form of the final spectrum, as illustrated in Fig. \ref{fig2}.

\begin{figure}[htb]
\vspace{-0.25cm}
\includegraphics[width=.5\textwidth]{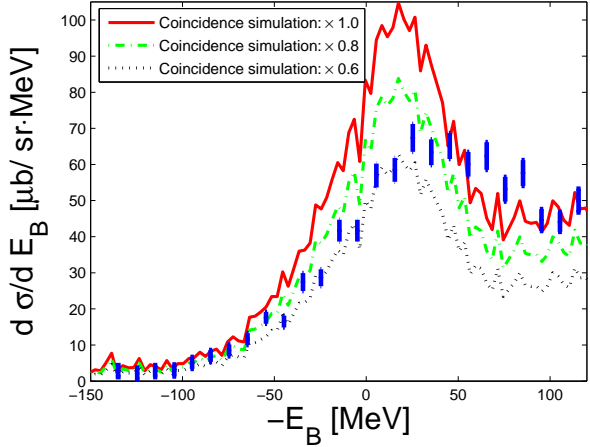}
\vspace{-0.75cm}
\caption{Calculated $ ^{12}C(K^-,p)$ spectrum 
 with  $V_{\rm opt}=(-60,-60)\rho/\rho_0$ MeV with minimal coincidence requirement - solid line; 
 and with additional suppression factors - dash-dotted and dotted lines. 
Experimental points are from \cite{Kishimoto:2007zz}.}
\label{fig3}
\end{figure}

To further simulate the coincidence requirement we introduce additional constant suppression factors 
to the obtained spectrum - see Fig. \ref{fig3}. Comparing our results with the experimental data we can 
conclude that in the "bound" region, $-E_B < 0$ MeV, these additional suppression 
is about $\sim 0.7$  and more or less homogeneous, while in the  continuum the suppression weakens and for  
$-E_B > 50$ MeV it is negligible. This picture is natural from the physical point of view, because the 
r.h.s. of the spectrum, Fig. \ref{fig3},  with relatively low momentum protons is mostly populated by many particle final states, 
which have a good chance to score the coincidence. 

To conclude, the main point of our analysis is not to state that
the data of Ref.~ \cite{Kishimoto:2007zz} supports  
${\rm Re}\, V_{\rm opt}=-60\rho/\rho_0$ MeV
rather than $-200\rho/\rho_0$. We want to make it clear that trying to simulate
these data one necessarily introduces large uncertainties due to the
experimental set up.   
Thus, this experiment is not appropriate
 for extracting information on the kaon optical potential. 

Contrary to what it is assumed in Ref.~\cite{Kishimoto:2007zz},
we clearly see, Fig. \ref{fig2}, that the spectrum shape is 
affected by the required coincidence. 
The experimental data without the coincidence requirement 
would be a more useful observable.

{\bf Acknowledgments.}\ \ \ 
This work is partly supported by
the contracts FIS2006-03438, FIS2008-01661 from MICINN
(Spain), by CSIC and JSPS under the Spain-Japan research Cooperative program,
 and by the Ge\-ne\-ra\-li\-tat de Catalunya contract 2009SGR-1289. We
acknowledge the support of the European Community-Research Infrastructure
Integrating Activity ``Study of Strongly Interacting Matter'' (HadronPhysics2,
Grant Agreement n. 227431) under the Seventh Framework Programme of EU.
J.Y. is a Yukawa Fellow and this work is partially supported by the
Yukawa Memorial Foundation.

\label{last}
\end{document}